\documentclass[prd,nofootinbib,superscriptaddress,tightenlines,notitlepage,twocolumn,floatfix]{revtex4-1}
\usepackage{bm}
\usepackage{tensor}
\usepackage{graphicx}
\usepackage{amssymb}
\usepackage{amsmath}
\usepackage{bm}
\usepackage{enumitem}  
\newcommand{\rmd}{{\rm d}}
\newcommand{\hata}{\hat{a}}
\newcommand{\hatb}{\hat{b}}
\newcommand{\hatc}{\hat{c}}
\newcommand{\abar}[1][]{{\left(\overline{a}_{\rm eff}^{#1}\right)}}
\newcommand{\anobar}[1][]{{\left(a_{\rm eff}^{#1}\right)}}

\begin{document}
\title{Lorentz-violating matter-gravity couplings in
small-eccentricity binary pulsars}\thanks{Invited article to special issue ``Symmetry in Special and General Relativity'' in MDPI journal {\it Symmetry}.}
\author{Lijing Shao}\email{Corresponding author: lshao@pku.edu.cn}
\affiliation{Kavli Institute for Astronomy and Astrophysics, Peking University, Beijing 100871, China}
\affiliation{Max-Planck-Institut f\"ur Radioastronomie, Auf dem H\"ugel 69,
D-53121 Bonn, Germany}
\date{\today}

\begin{abstract}
    Lorentz symmetry is an important concept in modern physics.
    Precision pulsar timing was used to put tight constraints on the
    coefficients for Lorentz violation in the pure-gravity sector of the
    Standard-Model Extension (SME). We extend the analysis to
    Lorentz-violating matter-gravity couplings, utilizing three
    small-eccentricity relativistic neutron star (NS) -- white dwarf (WD)
    binaries. We obtain compelling limits on various SME coefficients
    related to the neutron, the proton, and the electron. These results are
    complementary to limits obtained from lunar laser ranging and clock
    experiments.
\end{abstract}
\maketitle

\section{Introduction} \label{sec:intro}

The theory of general relativity (GR) and the Standard Model (SM) of
particle physics represent our contemporary condensed wisdom in the search
of fundamental laws in physics. Nevertheless, there exist various
motivations to look for new physics. Among them, the possibility of Lorentz
violation is a well developed concept~\cite{Tasson:2014dfa}. Lorentz
violation could be resulted from a deep underlying theory of quantum
gravity~\cite{Kostelecky:1988zi}. At low energy, it is believed to be
described by an effective field theory (EFT). An EFT framework, the
so-called Standard-Model Extension (SME), systematically incorporates all
Lorentz-covariant, gauge-invariant, energy-momentum-conserving operators
that are associated with GR and SM fields~\cite{Colladay:1996iz,
Colladay:1998fq, Kostelecky:2003fs}. Field operators are sorted according
to their mass dimension, and for some certain species, operators of
arbitrary mass dimensions are classified~\cite{Kostelecky:2009zp,
Kostelecky:2011gq, Kostelecky:2013rta, Kostelecky:2018yfa}.

The SME is supposed to be an effectively low-energy theory for the quantum
gravity, thus the gravitational aspect of the SME is of particular
interests. \citet{Kostelecky:2003fs} presented the general structure of the
SME when the curved spacetime is considered. \citet{Bailey:2006fd} worked
out different kinds of observational phenomena associated with the minimal
operators in the pure-gravity sector of the SME whose mass dimension $d
\leq 4$. After that, \citet{Kostelecky:2010ze} investigated in great detail
the theoretical aspects of the matter-gravity couplings whose mass
dimension $d \leq 4$. Phenomenological aspects and relevant experiments are
identified. Moreover, the nonminimal SME with gravitational operators whose
mass dimension $d > 4$ was studied and gained global interests during the
past few years~\cite{Bailey:2014bta, Shao:2016cjk, Kostelecky:2017zob}.

Due to the advances on the theoretical side~\cite{Kostelecky:2003fs,
Bailey:2006fd, Kostelecky:2010ze, Bailey:2014bta}, phenomenological and
experimental studies of the gravitational SME became a hot
topic~\cite{Tasson:2016blg, Bailey:2019rjj, Shao:2019nso, Tasson:2019kuw}.
\citet{Hees:2016lyw} have a comprehensive summary on this topic; see also
the {\it Data Tables for Lorentz and CPT Violation}, compiled by
\citet{Kostelecky:2008ts}. In the pure-gravity sector, binary pulsars turn
out to be among the best experiments in constraining, (i) the $d\leq 4$
minimal Lorentz-violating operators~\cite{Shao:2014oha, Shao:2014bfa}, (ii)
dimension-5 CPT-violating operators~\cite{Shao:2018vul}, as well as (iii)
dimension-8 cubic-in-the-Riemannian-tensor operators which are related to
the leading-order violation of the gravitational weak equivalence
principle~\cite{Shao:2019cyt}. In a closely related metric-based framework,
the so-called parameterized post-Newtonian formalism~\cite{Will:2014kxa,
Will:2018bme}, similarly, binary pulsars outperform many Solar-system-based
experiments~\cite{Shao:2012eg, Shao:2013wga, Shao:2013eka, Shao:2016ezh}.

In this work, we investigate the matter-gravity couplings in the SME and
their signals in binary pulsars~\cite{Kostelecky:2010ze, Jennings:2015vma}.
In particular, we use small-eccentricity binary pulsars,
PSRs~J0348+0432~\cite{Antoniadis:2013pzd},
J0751+1807~\cite{Desvignes:2016yex}, and J1738+0333~\cite{Freire:2012mg},
to put stringent constraints on various matter-gravity coupling
coefficients. The limits are compelling, and complementary to other
experiments. They contribute to the research field on the experimental
examination of the SME.

The paper is organized as follows. In the next section, we review the
matter-gravity couplings in the SME~\cite{Kostelecky:2010ze}. Then in
Sec.~\ref{sec:psr}, the orbital dynamics for a binary
pulsar~\cite{Jennings:2015vma} is provided. In particular, the secular
change of the eccentricity vector (decomposed into the two Laplace-Lagrange
parameters~\cite{Lange:2001rn}), and the secular change of the pulsar's
projected semimajor axis are discussed. Constraints on the matter-gravity
coupling coefficients are given in Sec.~\ref{sec:res}. The last section
discusses constraints from other experiments, the strong-field aspects of
pulsars, and the prospects in improving the limits on the Lorentz-violating
matter-gravity couplings.

\section{Matter-gravity couplings in the SME} \label{sec:sme}

In order to incorporate fermion-gravity couplings, we use the vierbein
formalism~\cite{Kostelecky:2003fs}. In the SME, the action for a massive
Dirac fermion $\psi$ reads~\cite{Kostelecky:2010ze},
\begin{align}
    \label{eq:Spsi}
    S_\psi = \int e\left( \frac{1}{2} i \tensor{e}{^\mu_a} \overline\psi
    \Gamma^a \overleftrightarrow{D}_\mu \psi - \overline{\psi} M \psi
    \right) {\rm d}^4 x \,,
\end{align}
where, for spin-independent cases,
\begin{align}
    \Gamma^a &\equiv \gamma^a - c_{\mu\nu} e^{\nu a}
    \tensor{e}{^\mu_b} \gamma^b - e_\mu e^{\mu a} \,, \\
    M &\equiv m + a_\mu \tensor{e}{^\mu_a} \gamma^a \,.
\end{align}
Here $\tensor{e}{_\mu^a}$ is the vierbein with $e$ its determinant; $m$ is
the mass of the fermion; $\gamma^a$ is the Dirac matrix; $a_\mu$,
$c_{\mu\nu}$, and $e_\mu$ are species-dependent, spin-independent
coefficient fields for Lorentz violation [see Eq.~(7) and Eq.~(8) in
Ref.~\cite{Kostelecky:2010ze} for spin-dependent terms].

While being kept to the leading order, a field redefinition via a
position-dependent component mixing in the spinor space can be used to show
that, the CPT-odd coefficients $a_\mu$ and $e_\mu$ always appear in the
combination~\cite{Kostelecky:2010ze},
\begin{equation}
    \anobar_\mu \equiv a_\mu - m e_\mu \,.
\end{equation}
Therefore, we shall consider only $\anobar_\mu$ and $c_{\mu\nu}$ in the
following.

At leading order, the point-particle action is~\cite{Kostelecky:2010ze},
\begin{align}\label{eq:Su}
    S_u = \int {\rm d}\lambda \left[ -m \sqrt{-\left( g_{\mu\nu} +
    2c_{\mu\nu} \right) u^\mu u^\nu} - \anobar_\mu u^\mu \right] \,,
\end{align}
where $u^\mu \equiv {\rm d}x^\mu / {\rm d}\lambda$. For a macroscopic
composite object, the action (\ref{eq:Su}) is still applicable with the
replacements~\cite{Kostelecky:2010ze},
\begin{align}
    m &\to \sum_{w} N^{w} m^{w} \,, \label{eq:composite:m} \\
    c_{\mu\nu} &\to \frac{\sum_{w} N^{w} m^{w}\left(c^{w}\right)_{\mu
    \nu}}{\sum_{w} N^{w} m^{w}} \,, \label{eq:composite:c} \\
    \anobar_{\mu} &\to \sum_{w} N^{w} {\anobar[w]}_{\mu}
    \label{eq:composite:a} \,,
\end{align}
where $w$ denotes the particle species, and $N^w$ is the number of
particles of type $w$. We have neglected the contribution from binding
energies which could be at most $\sim20\%$ for neutron stars (NSs), unless
some unknown nonperturbative effects take place (see discussions in
Sec.~\ref{sec:disc})~\cite{Shao:2016ezh}. In general the role of binding
energy could further aid the analysis of signals for Lorentz violation; see
Sec.~VI\,B in Ref.~\cite{Kostelecky:2010ze} for more details. Hereafter,
for simplicity we only consider three types of fermions, (i) the electron
$w=e$, (ii) the proton $w=p$, and (iii) the neutron $w=n$. In
Table~\ref{tab:composite}, we list the estimated composition of these three
species for NSs and white dwarfs (WDs), and their corresponding composite
coefficient fields for Lorentz violation.

\begin{table*}
    \caption{Estimated composition for NSs and WDs. Composite coefficient
    fields for Lorentz violation are estimated according to
    Eqs.~(\ref{eq:composite:m}--\ref{eq:composite:a}). In the table,
    $M_{\rm NS}$ and $M_{\rm WD}$ are the masses for NS and WD,
    respectively, and $m^n$ ($\simeq m^p$) is the mass for a neutron
    (proton) particle. We define $N_{\rm NS} \equiv M_{\rm NS}/m^n$ and
    $N_{\rm WD} \equiv M_{\rm WD}/m^n$.
    \label{tab:composite}}
    \centering
    \def\arraystretch{1.5}
    \begin{tabular}{lcc}
    \hline\hline
     & \textbf{Neutron Stars} & \textbf{White Dwarfs}	\\
    \hline
    Electron number, $N^e$ & $\sim 0$ & $\frac{1}{2} N_{\rm WD}$ \\
    Proton number, $N^p$ & $\sim 0$ & $\frac{1}{2} N_{\rm WD}$ \\
    Neutron number, $N^n$ & $N_{\rm NS}$ & $\frac{1}{2} N_{\rm WD}$ \\
    Composite $m$ & $M_{\rm NS}$ & $M_{\rm WD}$ \\
    Composite $c_{\mu\nu}$ & $c_{\mu\nu}^n$ & $ \frac{1}{2}\left(
    c^n_{\mu\nu} + c^p_{\mu\nu} + 0.0005 \, c^e_{\mu\nu}
    \right)$ \\
    Composite $\anobar_\mu$ & $N_{\rm NS} \anobar[n]_\mu $ &
    $\frac{1}{2} N_{\rm WD} \left[ \anobar[n]_\mu + \anobar[p]_\mu
    + \anobar[e]_\mu \right]$ \\
    \hline
    \end{tabular}
\end{table*}

In general, the coefficient fields, $\anobar_\mu$ and $c_{\mu\nu}$, are
dynamical fields. In the Riemann-Cartan spacetime, the Lorentz violation
often needs to be {\it spontaneous}~\cite{Bluhm:2019ato},
instead of {\it
explicit}~\cite{Kostelecky:2003fs}. The coefficient fields obtain their
vacuum expectation values via the Higgs-like spontaneous symmetry breaking
mechanism. We denote the vacuum expectation values of $\anobar_\mu$ and
$c_{\mu\nu}$, as $\abar_\mu$ and $\overline c_{\mu\nu}$, respectively. The
barred quantities are also known as the {\it coefficients for Lorentz
violation}~\cite{Kostelecky:2008ts}. In asymptotically inertial Cartesian
coordinates, they are assumed to be small and
satisfy~\cite{Kostelecky:2010ze},
\begin{align}
    \partial_\alpha \abar_\mu &= 0 \,, \\
    \partial_\alpha \overline c_{\mu\nu} &= 0 \,.
\end{align}
The coefficients for Lorentz violation, $\abar_\mu$ and $\overline
c_{\mu\nu}$~\cite{Kostelecky:2008ts}, are the quantities that we want to
investigate with pulsar timing experiments~\cite{Lorimer:2005misc,
Wex:2014nva} in this work.

\section{Binary pulsars with Lorentz-violating matter-gravity couplings} 
\label{sec:psr}

\citet{Jennings:2015vma} worked out the osculating elements for a binary
system, composed of masses $M_1$ and $M_2$, in the presence of the
Lorentz-violating matter-gravity couplings. We consistently use the
subscript ``1'' to denote the pulsar, and use the subscript ``2'' to denote
the companion which is a WD in our study. We define $q \equiv M_1/M_2$ and
$M \equiv M_1 + M_2$. To simplify some expressions, we also define $X
\equiv M_1 / M = q/\left( 1+q\right)$; then $M_2/M = 1-X =
1/\left(1+q\right)$.

\begin{figure}
    \centering
    \includegraphics[width=8cm]{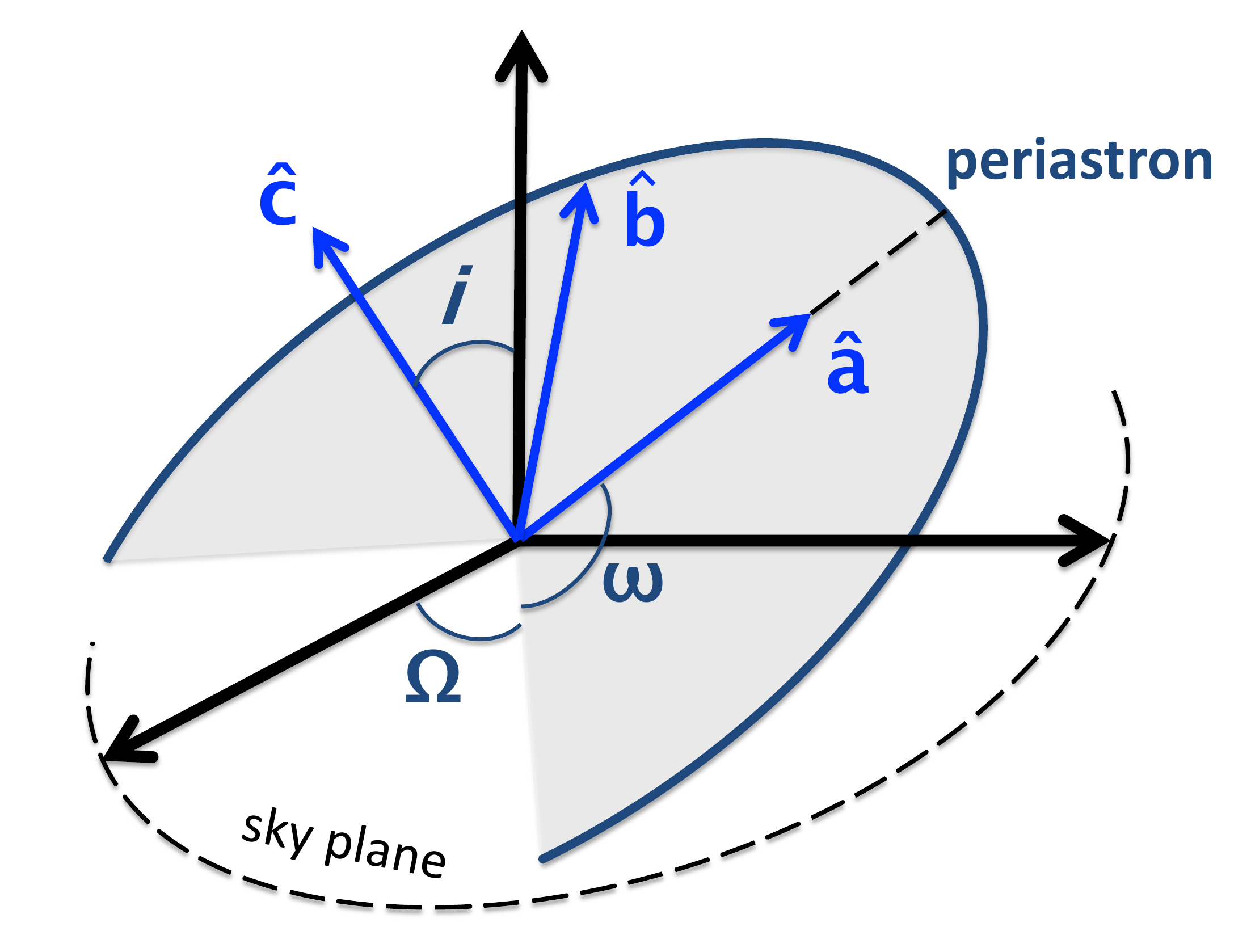} \caption{Pulsar orbit and
    the coordinate system $\left({\bf \hata}, {\bf \hatb}, {\bf \hatc}
    \right)$~\cite{Bailey:2006fd, Shao:2014bfa, Shao:2018vul}.
    \label{fig:geometry}}
\end{figure}

Neglecting the finite-size effects, the Newtonian relative acceleration for
a binary is $\bm{a}_{\rm N} = - GM_1 M_2/r^2 \hat{\bm{r}}$, where $\bm{r}$
is the relative separation and $\hat{\bm{r}} \equiv \bm{r}/r$. In the
Newtonian gravity, a two-body system with a negative total orbital energy
forms an elliptical orbit. An elliptical orbit in the celestial mechanics
is usually described by six orbital elements, (i) the semimajor axis $a$,
(ii) the orbital eccentricity $e$, (iii) the epoch of periastron passage
$T_0$, (iv) the inclination of orbit $i$, (v) the longitude of periastron
$\omega$, and (vi) the longitude of ascending node $\Omega$. The last three
angles are illustrated in Figure~\ref{fig:geometry}.

When there is a perturbing acceleration to $\bm{a}_{\rm N}$, say, $\delta
\bm{a}$, the orbit is changed perturbatively. In the osculating-element
approach, one assumes that at any instant moment, the orbit is still an
ellipse, but the six orbital elements become functions of the time
$t$~\cite{Poisson:2014misc}. The time derivatives of these six functions
are derived from the extra acceleration $\delta
\bm{a}$~\cite{Poisson:2014misc}. In the current case, after averaging over
an orbital period $P_b$, the secular changes read~\cite{Jennings:2015vma},
\begin{align}
    \left\langle\frac{\rmd a}{\rmd t}\right\rangle &= 0 \,, \label{eq:dadt} \\
    \left\langle\frac{\rmd e}{\rmd t}\right\rangle
    &=\frac{n_b}{M} \gamma \left(\frac{e^{2}-2
    \varepsilon}{e^{3}} A_{\hata \hatb}+\frac{n_b a \varepsilon}{e^{2}}
    B_{\hata}\right) \,, \label{eq:dedt} \\
    \left\langle\frac{\rmd i}{\rmd t}\right\rangle
    &=\frac{n_b}{M\gamma} \times \nonumber \\ 
    & \left(\frac{\varepsilon}{e^{2}} A_{\hata \hatc} \cos
    \omega-\frac{e^{2}-\varepsilon}{e^{2}} A_{\hatb \hatc} \sin
    \omega-\frac{n_b \varepsilon a}{e} B_{\hatc} \sin \omega\right) \,,
    \label{eq:didt} \\
    \left\langle\frac{\rmd \omega}{\rmd t}\right\rangle &=-\frac{n_b}{M
    \gamma \tan i}  \times \nonumber \\ 
    & \left(\frac{\varepsilon}{e^{2}} A_{\hata \hatc} \sin
    \omega+\frac{e^{2}-\varepsilon}{e^{2}} A_{\hatb\hatc} \cos
    \omega+\frac{n_b \varepsilon a}{e} B_{\hatc} \cos \omega\right) \nonumber \\
    &+\frac{n_b}{M}\left[\frac{e^{2}-2 \varepsilon}{2
    e^{4}}\left(A_{\hatb\hatb}-A_{\hata\hata}\right)+\frac{n_b
    a\left(1-\gamma\right)}{e^{3}} B_{\hatb}\right] \,, \label{eq:domdt}
\end{align}
where we have defined $\gamma \equiv \sqrt{1-e^2}$, $\varepsilon \equiv 1-
\gamma = 1- \sqrt{1-e^2}$, and $n_b \equiv 2\pi/P_b$. From
Eq.~(\ref{eq:dadt}), we can see that the energy of the orbit is conserved
at leading order, which is compatible with the action formulation of the
system in the absence of gravitational waves. The expressions for
$\left\langle \rmd \Omega / \rmd t \right\rangle$ and $\left\langle \rmd
T_0 / \rmd t \right\rangle$ are not important in the present context, thus
not shown. The 3-vector $B_j$ and the $3\times3$ tensor $A_{jl}$ are
defined as~\cite{Jennings:2015vma},
\begin{align}
    A_{jl} &=\sum_{w} 2 n_{7}^{w} m^{w} \overline{c}_{(j l)}^{w} \,, \\
    B_{j} &=-\sum_{w} 2 \left[n_{2}^{w}
    \left(\overline{a}_{\mathrm{eff}}^{w}\right)_{j} +\left(n_{6}^{w}-2
    n_{8}^{w}\right) m^{w} \overline{c}_{(0 j)}^{w} \right] \,,
\end{align}
where $n_i^w$ $(i=1,\cdots,8)$ are defined in Eq.~(9) of
Ref.~\cite{Jennings:2015vma}, and their approximated values for NS-NS and
NS-WD binaries are given in Table~\ref{tab:n} for convenience.

\begin{table*}
    \caption{Expressions of $n_i^w/N$ ($i=1,\cdots,8$;
    $w\in\left\{n,p,e\right\}$) for NS-NS and NS-WD systems
    (see Eq.~(9) in Ref.~\cite{Jennings:2015vma}), where $N \equiv
    N_1 + N_2 \simeq M/m^n$. Results in Table~\ref{tab:composite}
    are adopted for the calculation here. \label{tab:n}}
    \centering
    \def\arraystretch{1.5}
    \begin{tabular}{lcccccc}
    \hline\hline
     & \multicolumn{3}{c}{\textbf{Neutron Star -- Neutron Star}} &
     \multicolumn{3}{c}{\textbf{Neutron Star -- White Dwarf}} \\
     & $n$ & $\quad p \quad$ & $\quad e \quad$ & $n$ & $p$ & $e$ \\
    \hline
    $n_1^w/N$ & $1$ & 0 & 0 & $\frac{1}{2}(1+X)$ & $\frac{1}{2}(1-X)$ & $\frac{1}{2}(1-X)$ \\
    $n_2^w/N$ & $2X-1$ & 0 & 0 & $\frac{1}{2}(3X-1)$ & $-\frac{1}{2}(1-X)$ & $-\frac{1}{2}(1-X)$ \\
    $n_3^w/N$ & $2$ & 0 & 0 & $\frac{3}{2}$ & $\frac{1}{2}$ & $\frac{1}{2}$ \\
    $n_4^w/N$ & 0 & 0 & 0 & $-\frac{1}{2}$ & $\frac{1}{2}$ & $\frac{1}{2}$ \\
    $n_5^w/N$ & $2X\left(1-X\right)$ & 0 & 0 & $\frac{3}{2} X(1-X)$ & $\frac{1}{2}X(1-X)$ & $\frac{1}{2}X(1-X)$ \\
    $n_6^w/N$ & 0 & 0 & 0 & $-\frac{1}{2}X(1-X)$ & $\frac{1}{2}X(1-X)$ & $\frac{1}{2}X(1-X)$ \\
    $n_7^w/N$ & $1$ & 0 & 0 & $1-\frac{1}{2}X$ & $\frac{1}{2}X$ & $\frac{1}{2}X$ \\
    $n_8^w/N$ & $1-2X$ & 0 & 0 & $\frac{1}{2}X^2 -2X + 1$ & $-\frac{1}{2}X^2$ & $-\frac{1}{2}X^2$ \\
    \hline
    \end{tabular}
\end{table*}

In the above two equations, only $n_i^w$ with $i=2,6,7,8$ are relevant.
Using the results in Table~\ref{tab:n}, we have,
\begin{align}
    \frac{A_{jl}}{M} =& (2-X) \overline{c}^n_{(jl)} + X \left[
    \overline{c}^p_{(jl)} + 0.0005 \overline{c}^e_{(jl)} \right] \,, \\
    \frac{B_j}{M} =& \frac{1-X}{m^n} \left[ \abar[p]_j + \abar[e]_j \right]
    + \frac{1-3X}{m^n} \abar[n]_j \nonumber \\ 
    & + \left(X^2 -7X + 4\right)
    \overline{c}^n_{(0j)} \nonumber \\ 
    & - X(1+X) \left[ \overline{c}^p_{(0j)} + 0.0005 \overline{c}^e_{(0j)} \right] \,.
\end{align}
We can easily obtain the following conclusion from the above two equations.
(I) The sensitivity to $\overline{c}^e_{(jl)}$ and $\overline{c}^e_{(0j)}$
[compared with $\overline{c}^p_{(jl)}$ and $\overline{c}^p_{(0j)}$
respectively] is suppressed by the mass ratio of the electron to the proton
$(m^e / m^p \simeq 0.0005)$, while the sensitivity to $\abar[e]_j$
[compared with $\abar[p]_j$] is not suppressed. (II) We have no sensitivity
to $\abar[w]_0$ nor $\overline{c}^w_{00}$ $\left(w \in \left\{n,p,e\right\}
\right)$ from binary pulsars in this simplified situation. This is similar
to the case of $\overline{s}_{00}$ (the time-time component of the
 Lorentz-violating field $\overline{s}_{\mu\nu}$) in the pure-gravity
 sector of the SME with dimension 4 operators~\cite{Bailey:2006fd,
 Shao:2014oha}; nevertheless, these coefficients can be probed with the
 help of the ``boost effect'' introduced by the systematic velocity of the
 binary ($v^{\rm sys}/c \sim 10^{-3}$) with respect to the Solar
 system~\cite{Shao:2014bfa}. We defer the investigation along this line to
 future studies.

In Eqs.~(\ref{eq:dadt}--\ref{eq:domdt}), $B_j$ and $A_{jl}$ are projected to
the coordinate system $\left({\bf \hata}, {\bf \hatb}, {\bf \hatc}
\right)$~\cite{Bailey:2006fd, Shao:2014bfa, Shao:2018vul} where ${\bf \hata}$
is the unit vector points from the center of binary towards the periastron,
${\bf \hatc}$ is the unit vector points along the orbital angular momentum,
and ${\bf \hatb} \equiv {\bf \hatc} \times {\bf \hata}$ (see
Figure~\ref{fig:geometry}).

We are interested in the small-eccentricity binaries. In the limiting case of
small eccentricity $e \to 0$, we have
\begin{align}
    \gamma &= 1 - \frac{1}{2} e^2 - \frac{1}{8} e^4 + {\cal O}
    \left(e^6\right) \,, \\
    \varepsilon &= \frac{1}{2} e^2 + \frac{1}{8} e^4 + {\cal O}
    \left(e^6\right) \,.
\end{align}
Therefore, Eqs.~(\ref{eq:dedt}--\ref{eq:domdt}) are simplified to,
\begin{align}
    \left\langle\frac{\rmd e}{\rmd t}\right\rangle &\simeq \frac{n_b^2 a}{2M}
    B_{\hata}  \,, \\
    \left\langle\frac{\rmd i}{\rmd t}\right\rangle &\simeq \frac{n_b}{2M} \left(
    A_{\hata\hatc} \cos\omega - A_{\hatb\hatc} \sin\omega \right)  \,, \\
    \left\langle\frac{\rmd \omega}{\rmd t}\right\rangle &\simeq \frac{n_b^2 a}{2eM} B_{\hatb} \,.
\end{align}
We can convert the derivatives of $e$, $i$, and $\omega$, into derivatives of
the projected semimajor axis of the pulsar orbit $x_p$, and the
Laplace-Lagrange parameters, $\eta \equiv e\sin\omega$ and $\kappa \equiv
e\cos\omega$,
\begin{align}
  \left\langle\frac{\rmd x_p}{\rmd t}\right\rangle &= \frac{M_2 \cos i}{2M^2}
  \left(GMn_b\right)^{1/3} \left( A_{\hata\hatc} \cos\omega - A_{\hatb\hatc}
  \sin\omega \right) \,, \label{eq:dxpdt} \\
  \left\langle\frac{\rmd \eta}{\rmd t}\right\rangle &= \frac{n_b}{2M}
  \left(GMn_b\right)^{1/3} \left( B_{\hata}\sin\omega + B_{\hatb}\cos\omega
  \right) \,, \label{eq:detadt}\\
  \left\langle\frac{\rmd \kappa}{\rmd t}\right\rangle &= \frac{n_b}{2M}
  \left(GMn_b\right)^{1/3} \left( B_{\hata}\cos\omega - B_{\hatb}\sin\omega
  \right) \,, \label{eq:dkappadt}
\end{align}
where we have used $n_b a = \left(GMn_b\right)^{1/3}$.

\section{Bounds on the SME coefficients} \label{sec:res}

We use the time derivatives of $x_p$, $\eta$, and $\kappa$ in
Eqs.~(\ref{eq:dxpdt}--\ref{eq:dkappadt}) to constrain the coefficients for
Lorentz violation. It is clear that the more relativistic the binary
(namely, the larger $n_b$), the better the tests. Therefore, we use three
well-timed NS-WD binaries whose orbital periods are shorter than half a
day~\cite{Antoniadis:2013pzd, Desvignes:2016yex, Freire:2012mg}. Relevant
parameters of these binaries are collected in Table~\ref{tab:psr}. Due to
the binary interaction and matter exchange in the evolutionary history,
these NS-WD binaries have small orbital eccentricity $e \leq 10^{-6}$, thus
Eqs.~(\ref{eq:dxpdt}--\ref{eq:dkappadt}) are sufficient to perform the
tests.

\begin{table*}
    \caption{Relevant parameters for PSRs J0348+0432~\cite{Antoniadis:2013pzd},
    J0751+1807~\cite{Desvignes:2016yex}, and J1738+0333~\cite{Freire:2012mg}.
    Parenthesized numbers represent the 1-$\sigma$ uncertainty in the last
    digit(s) quoted. The parameter $\eta$ is the intrinsic value, after
    subtraction of the contribution from the Shapiro delay~\cite{Lange:2001rn}.
    Masses are derived without using information related to $\left\langle \rmd
    x_p / \rmd t \right\rangle$, $\left\langle \rmd \eta / \rmd t
    \right\rangle$, nor $\left\langle \rmd \kappa / \rmd t \right\rangle$
    for consistency. For PSRs~J0348+0432 and J1738+0333, masses were
    derived independently of gravity theories~\cite{Freire:2012mg,
    Antoniadis:2013pzd}, while for PSR~J0751+1807 we have used observed
    quantities related to the Shapiro delay and orbital decay, assuming the
    validity of GR~\cite{Desvignes:2016yex}.
    \label{tab:psr}}
    \centering
    \def\arraystretch{1.5}
    \begin{tabular}{lccc}
    \hline\hline
    \textbf{Pulsar} & \textbf{J0348+0432} & \textbf{J0751+1807} & \textbf{J1738+0333} 	\\
    \hline
    Observational span, $T_{\rm obs}$\,(year) & $\sim3.7$ & $\sim17.6$ & $\sim10.0$ \\
    Orbital period, $P_b$\,(day) & 0.102424062722(7) & 0.263144270792(7) & 0.3547907398724(13) \\
    Pulsar's projected semimajor axis, $x_p$\,(lt-s) & 0.14097938(7) & 0.3966158(3) & 0.343429130(17) \\
    $\eta \equiv e\sin \omega\,(10^{-7})$ & 19(10) &  33(5) & $-1.4(11)$ \\
    $\kappa \equiv e\cos \omega\,(10^{-7})$ & 14(10) & 3.8(50) & 3.1(11) \\
    Time derivative of $x_p$, $\dot x_p$ & -- & $(-4.9\pm0.9)\times10^{-15}$ &  $(0.7\pm0.5)\times10^{-15}$ \\
    NS mass, $m_1\,(M_\odot)$ & 2.01(4) & 1.64(15) & $1.46^{+0.06}_{-0.05}$ \\
    WD mass, $m_2\,(M_\odot)$ & 0.172(3) & 0.16(1) &  $0.181^{+0.008}_{-0.007}$ \\
    \hline
    \end{tabular}
\end{table*}

From Table~\ref{tab:psr}, we see that the time derivatives of $\eta$ and
$\kappa$ are not reported in literature, as well as the time derivative of
$x_p$ for PSR~J0348+0432. The reason is usually the following. In fitting
the times of arrival of pulse signals, these quantities would be measured
to be consistent with zero if they were included in the timing formalism.
To have a simpler timing model, these quantities are considered {\it
unnecessary} for a good fit. Actually, the insignificance of these
quantities is consistent with the spirit of our tests to put upper limits
on the Lorentz violation. We estimate the upper limits for these quantities
using $\dot X \sim \sqrt{12} \sigma_X / {T_{\rm obs}} $ $(X \in \left\{x_p,
\eta, \kappa\right\})$~\cite{Shao:2014oha}, where $\sigma_X$ is the
measured uncertainty for the quantity $X$ and $T_{\rm obs}$ is the
observational span of the data from where these quantities were derived.
The factor ``$\sqrt{12}$'' approximately takes a linear-in-time evolution
of the quantity $X$ into account~\cite{Shao:2014oha}. It is verified that
this approximation works reasonably well~\cite{Shao:2014oha, Shao:2018vul}.
For PSRs~J0751+1807 and J1738+0333, $\left\langle \rmd x_p / \rmd t
\right\rangle$ was measured to be nonzero. Because the proper motion of the
binary in the sky could contribute to a nonzero $\left\langle \rmd x_p /
\rmd t \right\rangle$ for nearby pulsars~\cite{Kopeikin:1996ads,
Lorimer:2005misc}, we use the measured value of $\left\langle \rmd x_p /
\rmd t \right\rangle$ as an upper limit for the effects from Lorentz
violation. For nearby pulsars, the contribution to $\left\langle \rmd
x_p / \rmd t \right\rangle$ from the proper motion depends sinusoidally on
$\Omega$~\cite{Kopeikin:1996ads, Lorimer:2005misc}; although $\Omega$ is
not measured, we do not expect the Nature's conspiracy in assigning certain
values of $\Omega$, case by case to different binary pulsars, in order to
hide the Lorentz symmetry breaking. Therefore, we believe the above
treatments introduce uncertainties no larger than a multiplicative factor
of a few.

In order to use Eqs.~(\ref{eq:dxpdt}--\ref{eq:dkappadt}), one also needs
the absolute geometry of the orbit to properly project the vector $B_j$ and
the tensor $A_{jl}$ onto the coordinate system $\left({\bf \hata}, {\bf
\hatb}, {\bf \hatc} \right)$. In general, the longitude of the ascending
node $\Omega$ is not an observable in pulsar
timing~\cite{Lorimer:2005misc}. Nevertheless, the procedure to randomize
the value of $\Omega \in [0,360^\circ)$ and to systematically project
vectors and tensors onto $\left({\bf \hata}, {\bf \hatb}, {\bf \hatc}
\right)$ was worked out in Ref.~\cite{Shao:2014oha}. It was successfully
applied to binary pulsars in previous studies~\cite{Shao:2014oha,
Shao:2014bfa, Shao:2018vul, Shao:2019cyt}. Since here (i) we have already
introduced an uncertainty with a factor of a few, and (ii) we are
interested in the ``maximal-reach'' limits in absence of the Lorentz
violation, we take a simplified approach and treat these projections as
${\cal O}(1)$ operators. The ``maximal-reach''
approach~\cite{Tasson:2019kuw} assumes that only one component of
Lorentz-violating coefficients is nonzero in a test. We think our approach
reasonable at the stage of setting upper limits to the coefficients for
Lorentz violation. Nevertheless, when people start to discover some
evidence for the Lorentz violation, it is {\it absolutely} needed to take
into account more sophisticated analysis, for example, to use the 3-D
orientation of the orbit (possibly in a probabilistic way with an unknown
$\Omega$) as was done in Refs.~\cite{Shao:2014oha, Shao:2014bfa,
Shao:2018vul, Shao:2019cyt}. In addition, if one wants to explore the
correlation between different coefficients for Lorentz violation, more
sophisticated analysis is needed as well. These improvements lay beyond the
scope of this work.

In Table~\ref{tab:res} we list the ``maximal-reach''~\cite{Tasson:2019kuw}
limits on the coefficients for Lorentz violation with matter-gravity
couplings obtained from binary pulsars. As we can see, the best limits on
$\overline{c}^w_{jk}$ $(w\in\left\{n,p,e\right\})$ come from PSR~J1738+0333
due to its very good measurement on the $\dot x_p$~\cite{Freire:2012mg}.
For $\overline{c}^w_{0k}$ and $\abar[w]_k$, the best limits come from
PSR~J0751+1807 due to its good measurement of the Lagrange-Laplace
parameters~\cite{Desvignes:2016yex}.

\begin{table*}
\caption{``Maximal-reach'' limits from binary pulsars on the coefficients for
Lorentz violation with matter-gravity couplings where, only one component
is assumed to be nonzero at a time. The limits on $\overline{c}^w_{jk}$
$(w\in\left\{n,p,e\right\})$ come from $\left\langle \rmd x_p / \rmd t
\right\rangle$, while the limits on $\overline{c}^w_{0k}$ and $\abar[w]_k$
come from $\left\langle \rmd \eta / \rmd t \right\rangle$ or $\left\langle
\rmd \kappa / \rmd t \right\rangle$, and only the stronger one is listed in
the table. For each row, the strongest limit is shown in boldface.
\label{tab:res}}
    \centering
    \def\arraystretch{1.5}
    \begin{tabular}{lccc}
    \hline\hline
    \textbf{SME Coefficients} & \textbf{PSR J0348+0432} & \textbf{PSR J0751+1807} & \textbf{PSR J1738+0333} 	\\
    \hline
    $\overline{c}^n_{jk}$ & $3\times10^{-11}$ & $2\times10^{-10}$ & $\bm{1\times10^{-11}}$ \\
    $\overline{c}^p_{jk}$ & $4\times10^{-11}$ & $2\times10^{-10}$ & $\bm{1\times10^{-11}}$ \\
    $\overline{c}^e_{jk}$ & $8\times10^{-8}$ & $4\times10^{-7}$ & $\bm{3\times10^{-8}}$ \\
    $\overline{c}^n_{0k}$ & $3\times10^{-8}$ & $\bm{1\times10^{-8}}$ & $7\times10^{-8}$ \\
    $\overline{c}^p_{0k}$ & $2\times10^{-8}$ & $\bm{1\times10^{-8}}$ & $6\times10^{-8}$ \\
    $\overline{c}^e_{0k}$ & $5\times10^{-5}$ & $\bm{2\times10^{-5}}$ & $1\times10^{-4}$ \\
    $\abar[n]_k$ & $2\times10^{-8}\,{\rm GeV}$ & $\bm{1\times10^{-8}\,{\rm GeV}}$ &  $6\times10^{-8}\,{\rm GeV}$ \\
    $\abar[p]_k$ & $5\times10^{-7}\,{\rm GeV}$ & $\bm{2\times10^{-7}\,{\rm GeV}}$ & $8\times10^{-7}\,{\rm GeV}$ \\
    $\abar[e]_k$ & $5\times10^{-7}\,{\rm GeV}$ & $\bm{2\times10^{-7}\,{\rm GeV}}$ & $8\times10^{-7}\,{\rm GeV}$ \\
    \hline
    \end{tabular}
\end{table*}

\section{Discussions} \label{sec:disc}

Besides the streamlined theoretical analysis, the maximal-reach limits in
Table~\ref{tab:res} are the main results of this paper. As far as we are
aware, \citet{Altschul:2006uu} was the first to put preliminary limits on
the SME neutron-sector coefficients with pulsar rotations. The pure-gravity
sector of the SME at different mass dimensions was systematically tested
with binary pulsars in Refs.~\cite{Shao:2014oha, Shao:2014bfa,
Shao:2018vul, Shao:2019cyt}. Early limits on $\abar[w]_k$ were given with
K/He magnetometer and torsion-strip balance~\cite{Tasson:2012nx,
Panjwani:2010oga}; but these limits, while constraining different linear combinations of the Lorentz violating coefficients, are rather weak. Later the
maximal-reach limits on $\abar[w]_k$ were obtained systematically with
superconducting gravimeters~\cite{Flowers:2016ctv} and lunar laser ranging
(LLR) experiments~\cite{Bourgoin:2017fpo}. The former got $\abar[w]_k \leq
{\cal O}\left( 10^{-5}\, {\rm GeV} \right)$, while the latter got
$\abar[w]_k \leq {\cal O}\left( 10^{-8}\, {\rm GeV} \right)$. 
Our best limits from PSR~J0751+1807 for the proton and
the electron are weaker than the LLR limits, while our limit for the
neutron is slightly better. There is also a limits from the observation of
gravitational waves, but being weaker than our limits by almost 30 orders
of magnitude~\cite{Schreck:2016qiz}. The limits on $\abar[w]_0$ were cast
by analyzing nuclear binding energy, Cs interferometer, torsion pendulum,
and weak equivalence principle experiments~\cite{Kostelecky:2008in,
Hohensee:2011wt, Hohensee:2013hra, Kostelecky:2010ze}. The analysis with
binary pulsars in this work could not bound these SME coefficients. The
limits on $\bar c^w_{\mu\nu}$ from other experiments (e.g. clock
experiments~\cite{Bars:2016mew}) are much better than the limits from
binary pulsars~\cite{Kostelecky:2008ts}. However, our limits are the best
ones from gravitational systems. In a short summary, our limits are
compelling, and being complementary to limits obtained from other
experiments.

In using the SME, we have assumed the validity of the EFT and the smallness
of the Lorentz violation. This is true for most ordinary objects. However,
we shall be aware of a caveat for NSs, because of the possible
nonperturbative behaviors which might be triggered by their strongly
self-gravitating nature~\cite{Wex:2014nva}. It was shown explicitly that,
in a class of scalar-tensor theories, highly nonlinear phenomena are
possible within NSs and they may result in large deviations from
GR~\cite{Damour:1993hw, Damour:1996ke}. Although the nonperturbative
behaviors were constrained tightly with binary pulsars and the binary
neutron star inspiral GW170817~\cite{Freire:2012mg, Shao:2017gwu,
Zhao:2019suc}, the possibility is not completely ruled out
yet~\cite{Yunes:2016jcc, Sathyaprakash:2019yqt, Carson:2019fxr}. With this
caveat in mind, conservatively speaking, the tests in this paper are
basically testing the strong-field counterparts of the weak-field SME
coefficients. Usually, when the strong-field effects are considered, the
constraints become even tighter. Therefore, we treat the limits here
conservative ones~\cite{Shao:2016ezh}.

The tests of Lorentz violation with binary pulsars improve with a longer
baseline for data~\cite{Shao:2014oha}. Specifically, even {\it
pessimistically} assuming no advance in the quality of binary-pulsar
observation for the future, the tests in
Eqs.~(\ref{eq:dxpdt}--\ref{eq:dkappadt}) improve as $T_{\rm obs}^{-1.5}$
where $T_{\rm obs}$ is the total observational span. In reality, the
quality of observation improves rapidly, especially with the newly built
and upcoming telescopes, like the Five-hundred-meter Aperture Spherical
Telescope (FAST), the MeerKAT telescope, and the Square Kilometre Array
(SKA)~\cite{Kramer:2004hd, Shao:2014wja, Bull:2018lat, Bailes:2018azh}.
Therefore, we expect better tests than the $T_{\rm obs}^{-1.5}$ scaling in
testing the Lorentz violation in the future.

\begin{acknowledgements}
We are grateful to Jay Tasson for the invitation and stimulating
discussions. We thank Norbert Wex for carefully reading the manuscript,
Adrien Bourgoin, Zhi Xiao and Rui Xu for communication. This work was supported by 
the Young Elite Scientists Sponsorship Program by the
China Association for Science and Technology (2018QNRC001), and partially
supported by the National Natural Science Foundation of China (11721303),
the Strategic Priority Research Program of the Chinese Academy of Sciences
through the Grant No. XDB23010200, and the European Research Council (ERC)
for the ERC Synergy Grant BlackHoleCam under Contract No. 610058.
\end{acknowledgements}

\bibliography{refs}

\end{document}